\documentclass[11pt, english]{smfart}
\def\today{20.11.06} 
\usepackage{amsmath,amsfonts,amsthm,amssymb,amscd}

\newcommand{\be}{\begin{equation}}
\newcommand{\ee}{\end{equation}}
\newcommand{\bs}{\begin{split}}
\newcommand{\es}{\end{split}}
\newcommand{\ba}{\begin{align}}
\newcommand{\ea}{\end{align}}
\newcommand{\basl}[1]{\begin{align}\begin{split}\label{#1}}
\newcommand{\bas}{\begin{align}\begin{split}}

\theoremstyle{plain} \newtheorem{theorem}{Theorem}[section]
\newtheorem{lemma}[theorem]{Lemma}
\newtheorem{proposition}[theorem]{Proposition}

\newcommand{\R}{{\mathbb R}}

\newcommand{\N}{{\mathbb N}}

\newcommand{\F}{{\mathcal F}}
\newcommand{\Fq}{{\mathcal F^{(Q)}}}
\newcommand{\C}{\mathbb{C}}
 
\newcommand{\psiq}{\Psi^{(Q)}}
\newcommand{\xib}{\bar \xi}
\newcommand{\Om}{\Omega}
\newcommand{\Q}{{(Q)}}
\newcommand{\bjx}{b_{j,\epsilon}(\xi)}
\newcommand{\bsjx}{b^\star_{j,\epsilon}(\xi)}

\newcommand{\bupx}{b_{1,+}(\xi)}
\newcommand{\bumx}{b_{1,-}(\xi)}
\newcommand{\ep}{\epsilon}
\newcommand{\om}{\omega}
\newcommand{\al}{\alpha}
\newcommand{\bet}{\beta}
\newcommand{\ga}{\gamma}
\newcommand{\la}{\lambda}
\newcommand{\Pq}{\Psi^{(Q)}}
\newcommand{\Pqp}{\Psi^{(Q')}}
\newcommand{\Ff}{\F_{{\mathrm fin}}}
\newcommand{\his}{H_{I,\sigma}}
\newcommand{\pon}{P_{\Om_{\mathrm{neut}}}}

\newcommand{\va}[1]{|#1|}

  \newcommand{\vag}{|g|}
  \newcommand{\Va}[1]{\left|#1\right|}
\newcommand{\norm}[1]{\|#1\|}

\numberwithin{equation}{section}

\begin{document}

\author{
Laurent AMOUR,
\quad Beno\^\i t GR\'EBERT\\
 and \\ Jean-Claude GUILLOT
}

\title{A mathematical model for the Fermi weak interactions}

\date{\today }

\begin{abstract}
    We consider a mathematical model of the Fermi theory of 
    weak interactions as patterned according to the well-known 
    current-current coupling of quantum electrodynamics.
    We focuss on the example of the decay of the muons into electrons, 
    positrons and neutrinos but other examples are considered in 
    the same way.
    We prove that the Hamiltonian describing this model has a ground state in 
    the fermionic Fock space for a sufficiently small coupling constant. Furthermore 
    we determine the absolutely continuous spectrum of the Hamiltonian and 
    by commutator estimates we prove that the spectrum is absolutely continuous 
    away from a small neighborhood of the thresholds of the free Hamiltonian.
    For all these results we do not use any infrared cutoff 
    or infrared regularization even if fermions with zero mass are 
    involved.

\end{abstract}

\frontmatter
\maketitle
\newpage
\mainmatter
\tableofcontents
\setcounter{section}{0}

\section{Introduction}

In this note we consider a mathematical model of the Fermi theory of 
weak interactions as patterned according to the well-known 
current-current coupling of quantum electrodynamics (see 
\cite{Gre,Wei}). The weak interaction processes are well described at 
low energy by the current-current coupling.\\
We choose the example of the decay of the muons into electrons, 
positrons and neutrinos.
The beta decay of the neutron  
could be considered too.

The mathematical framework involves a fermionic Fock space for the 
particles and the antiparticles and the interaction is described in 
terms of annihilation and creation operators together with an 
$L^2$-kernel with respect to the momenta. The total Hamiltonian, 
which is the sum of the free energy of the particles and the 
antiparticles and of the interaction, is a self-adjoint operator in 
the Fock space. We prove that this Hamiltonian has a ground state in 
the Fock space for a sufficiently small coupling constant. Furthermore 
we determine the absolutely continuous spectrum of the Hamiltonian and 
by commutator estimates we prove that the spectrum is absolutely continuous 
    away from a small neighborhood of the thresholds of the free Hamiltonian.\\
From the mathematical point of view, the interaction is no more 
invariant by translation and the singularity of the kernel at the 
origin is not too strong. In fact the physical formal kernel is 
locally bounded at the origin. This means that there is no infrared 
problem even if fermions with zero mass are involved in the model in 
contrast to the case of QED. Detailed proofs are only given for the 
Hamiltonian associated with the decay of muons.\\
We also describe the mathematical model for the beta dacay of quarks 
$u$ and $d$ for which the results will be the same. We also consider 
the decay of the massive bosons $W^+$ and $W^-$.

For the proofs we essentially follow the methods developed in 
\cite{BFS98b} \cite{BDG} and in \cite{AGG06b} for the existence of the ground state 
and those developed by \cite{BFS98b} and \cite{Sk} for the study of the 
continuous singular spectrum.

Let us finally mention that the same results should hold in Fock spaces 
associated to the Dirac equation in Schwarschild, Reisner-Nordstr{\o}m 
and Kerr black holes as soon as a generalized eigenfunction expansion 
for the Dirac equation in that context is known.


\section{The model}

The decay of the muons involves four species of particles and 
antiparticles, the muons $\mu^{-}$ and $\mu^{+}$, the electron $e^{-}$ 
and the positron $e^{+}$, the neutrino $\nu_{e}$ and the antineutrino 
$\bar\nu_{e}$ associated to the electron and the neutrino $\nu_{\mu}$
 and the  antineutrino $\bar\nu_{\mu}$ associated to the muon.\\
 In this article we consider the neutrinos $\nu_{e}$ and $\nu_{\mu}$ 
together with the antineutrinos $\bar\nu_{e}$ and $\bar\nu_{\mu}$ 
as neutrinos and antineutrinos with different quantum leptonic 
numbers (see \cite{Gre}, \cite{Pesk}). Thus, according to the 
convention described in section 4.1 of \cite{We} and from the 
mathematical point of view, in what follows the corresponding 
creation and annihilation operators for $\nu_{e}$ and $\bar\nu_{e}$ 
will anticommute with those for $\nu_{\mu}$ and $\bar\nu_{\mu}$. Our 
proof does not work if the neutrinos $\nu_{e}$ and $\nu_{\mu}$ are 
considered as particles of different species i.e., if the corresponding 
creation and annihilation operators for $\nu_{e}$ and $\bar\nu_{e}$ 
commute with those for $\nu_{\mu}$ and $\bar\nu_{\mu}$. \\
Concerning our notations from now on the particles and antiparticles 
$1$ will be the electrons $e^{-}$ and the positrons $e^{+}$, the 
particles and antiparticles $2$ will be the neutrinos $\nu_{e}$, 
$\bar\nu_{e}$, the particles and antiparticles $3$ will be the 
neutrinos $\nu_{\mu}$, $\bar\nu_{\mu}$ and, finally, the particles and 
antiparticles $4$ will be the muons $\mu^{-}$ and $\mu^{+}$.\\
Let $\xi =(p,s)$ be the quantum variables of a particle of spin 
$1/2$. Here $p\in \R^3$ is 
the momentum, $s\in \{-1/2 ,1/2\}$ is the 
spin polarization of particles and antiparticles 1 and 4 and
$s\in \{-1 ,1\}$ is the 
helicity of particles and antiparticles 2 and 3.
We set $\Sigma_{1}=\R^3\times\{-1/2 
,1/2\}$ for the particles and antiparticles 1 and 4 and $\Sigma_{2}=
\R^3\times\{-1,1\}$ for
particles and antiparticles 2 and 3. We will denote 
by $\bar \xi$ the quantum variables of an antiparticle.\\
Let us define the Fock space. Set 
$$
Q=(q,\bar q,r,\bar r, s, \bar s, t, \bar t)\in \N^8$$
where $q$ (resp. $r,s,t$) is the number of particles 1 (resp. 2,3,4) 
and $\bar q$ (resp. $\bar r, \bar s, \bar t)$ is the number of 
antiparticles 1 (resp. 2,3,4). For $i=q,r,s,t$ and $\bar i= \bar q, 
\bar r, \bar s, \bar t$ we introduce the following sets of variables:
$$
\Xi_{i}=(\xi_{1},\xi_{2},\ldots,\xi_{i})       \quad \Xi_{\bar i}=
(\bar\xi_{1},\bar\xi_{2},\ldots,\bar\xi_{i}) .
$$
Notice that for the neutrinos and antineutrinos we could use another 
sets of variables by adding leptonic quantum numbers to the $\xi$'s 
in order to get an equivalent framework.\\
Let us denote by $\Psi^{(Q)}(\cdot)$ a measurable function of the set 
of variables $\Xi_{q},\Xi_{\bar q},\ldots,\Xi_{t},\Xi_{\bar t}$ which 
is antisymmetric with respect to each set of variables $\Xi_{i}$ and 
$\Xi_{\bar i}$ separately and which is square integrable:
$$
\norm{\Psi^{(Q)}}^2=\int \Va{\Psi^{(Q)}(\Xi_{q},\Xi_{\bar q},
\Xi_{r},\Xi_{\bar r},\Xi_{s},\Xi_{\bar s},\Xi_{t},\Xi_{\bar 
t})}^2\prod_{i=(q,r,s,t)}d\Xi_{i}\prod_{\bar i=(\bar q,\bar r,\bar s, 
\bar t)}d\Xi_{\bar i} < \infty$$
where $d\Xi_{i}=\prod_{k=1}^id\xi_{k}$, $d\xi=\sum_{s}\int d^3 p$ and 
$d\Xi_{\bar i}=\prod_{k=1}^id\bar\xi_{k}$, $d\bar\xi=\sum_{s}\int d^3 
\bar p$. When $i=0$ or $\bar i=0$, the corresponding variables do not 
appear in $\Psi^{(Q)}$.\\
The space $\Fq=\{\psiq \mid \norm{\psiq}< \infty \}$ is an Hilbert 
space and the Fock space is defined by
$$
\F= \oplus_{Q\in \N^8}\Fq$$
where $\F^{(0)}=\C$. The vacuum $\Om$ is the state $(\psiq)_{Q}$
with $\psiq =0 $ for $Q\neq 0$ and $\Psi^{(0)}=1$. $\F$ is an Hilbert 
space and if $\Psi=(\psiq)_{Q}\in \F$ we have
$$
\norm{\Psi}^2=\sum_{Q\in \N^8}\norm{\psiq}^2.$$
We can now define the formal annihilation and creation operators 
$\bjx$ and $\bsjx$ for each type of particles and antiparticles. We 
have 
\bas
(\bupx\Psi)^\Q&(\xi_{1},\ldots,\xi_{q};\Xi_{\bar q};
\Xi_{r};\Xi_{\bar r};\Xi_{s};\Xi_{\bar s};\Xi_{t};\Xi_{\bar 
t})
=\\ 
&\sqrt{q+1}\Psi^{(q+1,\bar q,\ldots,\bar t)}
(\xi,\xi_{1},\ldots,\xi_{q};\Xi_{\bar q};
\Xi_{r};\Xi_{\bar r};\Xi_{s};\Xi_{\bar s};\Xi_{t};\Xi_{\bar 
t})
\end{split}\end{align}
and
\bas
(\bumx\Psi)^\Q&(\Xi_{q};\xib_{1},\ldots,\xib_{\bar q};\Xi_{r};\ldots;\Xi_{\bar t})
=\\
&\sqrt{\bar q+1}(-1)^q\Psi^{(q,\bar q +1,\ldots,\bar t)}
(\Xi_{q};\xi,\xib_{1},\ldots,\xib_{\bar q};\Xi_{r};\ldots;\Xi_{\bar t}).
\end{split}\end{align}
The operators $b_{2,\pm}(\xi) $ (resp. $b_{4, \pm}(\xi)$) are defined 
similarly by substituting $r$ and $\bar r$ (resp. $t$ and $\bar t$) 
for $q$ and $\bar q$ in an obvious way.\\
Furthermore, taking into account the anticommutation between 
$b_{3,\pm}$ and $b_{2,\pm}$, we have
\bas
(b_{3,+}&(\xi)\Psi)^\Q(\Xi_{q};\Xi_{\bar q};
\Xi_{r};\Xi_{\bar r};\xi_{1},\ldots,\xi_{s};\Xi_{\bar s};\Xi_{t};\Xi_{\bar 
t}) 
=\\& \sqrt{s+1} (-1)^{r+\bar r}\Psi^{(q,\bar q, r, \bar r, s+1, \bar s, 
t, \bar t)}(\Xi_{q};\Xi_{\bar q};
\Xi_{r};\Xi_{\bar r};\xi,\xi_{1},\ldots,\xi_{s};\Xi_{\bar s};\Xi_{t};\Xi_{\bar 
t}) \end{split}\end{align}
and
\bas
(b_{3,-}&(\xi)\Psi)^\Q(\Xi_{q};\Xi_{\bar q};
\Xi_{r};\Xi_{\bar r};\Xi_{s};\xib_{1},\ldots,\xib_{\bar s};\Xi_{t};\Xi_{\bar 
t}) 
=\\& \sqrt{\bar s+1} (-1)^{r+\bar r+s}\Psi^{(q,\bar q, r, \bar r, s, \bar s +1, 
t, \bar t)}(\Xi_{q};\Xi_{\bar q};
\Xi_{r};\Xi_{\bar r};\Xi_{s};\xi,\xib_{1},\ldots,\xib_{\bar s};\Xi_{t};\Xi_{\bar 
t}). \end{split}\end{align}
As usual $\bsjx$ is the formal adjoint of $\bjx$, for example
\bas
(b^\star_{1,+}&(\xi)\Psi)^{(q+1,\bar q, r, \bar r, s, \bar s, t, \bar t)}
(\xi_{1},\ldots,\xi_{q+1};\Xi_{\bar q};
\Xi_{r};\Xi_{\bar r};\Xi_{s};\Xi_{\bar s};\Xi_{t};\Xi_{\bar 
t})=\\&
\frac{ 1}{ \sqrt{q+1}}\sum_{i=1}^{q+1} (-1)^{i+1}\delta(\xi -\xi_{i})
\Psi^{(q,\bar q, r, \bar r, s, \bar s, t, \bar t)}
(\xi_{1},\ldots,\hat\xi_{i},\ldots,\xi_{q+1};\Xi_{\bar q};
\Xi_{r};\Xi_{\bar r};\Xi_{s};\Xi_{\bar s};\Xi_{t};\Xi_{\bar 
t})
\end{split}\end{align}
where $\hat\cdot$ denotes that the ith variable has to be omitted.\\
The following canonical anticommutation relations hold
$$\{\bjx,b^\star_{j,\epsilon'}(\xi')\}=\delta_{\epsilon,\epsilon'}\delta(\xi-\xi'),
\quad 
j=1,2,3,4,\ \epsilon,\epsilon'=\pm$$
where $\delta(\xi-\xi')=\delta_{s,s'}\delta(p-p')$,
$$\{\bjx,b_{j,\epsilon'}(\xi')\}=
\{b^\star_{j,\epsilon}(\xi),b^\star_{j,\epsilon'}(\xi')\}=0, \quad j=1,2,3,4,\ 
\epsilon,\epsilon'=\pm$$
$$
\{b_{2,\epsilon}(\xi),b^\sharp_{3,\epsilon'}(\xi')\}=
\{b^\star_{2,\epsilon}(\xi),b^\sharp_{3,\epsilon'}(\xi')\}=0$$
where $b^\sharp$ is $b$ or $b^\star$.\\ Note that
$$[b_{i,\epsilon}(\xi),b^\sharp_{j,\epsilon'}(\xi')]=
[b^\star_{i,\epsilon}(\xi),b^\sharp_{j,\epsilon'}(\xi')]=0, \quad
j=1,2,3,4,\ i=1,4 \mbox{ and }j\neq i.$$
Let $\F_{0}$ be the subspace of functions $\Psi=(\psiq)_{Q}$ such 
that $\psiq$ is a function in the Schwartz space and $\psiq =0 $ for 
all but finitely many $Q$. The $\bjx$'s are well defined operators 
on $\F_{0}$ but they are not closable. It is better to introduce the 
following operators:
$$
b_{j,\epsilon}(\phi)=\int \bjx \overline{\phi(\xi)}d\xi,
$$
$$
b^\star_{j,\epsilon}(\phi)=\int \bsjx {\phi(\xi)}d\xi
$$
where $\phi\in L^2(\Sigma)$ and $\Sigma = \Sigma_{1}$ when $j=1,4$ 
and $\Sigma=\Sigma_{2}$ when $j=2,3$. Both $b_{j,\epsilon}(\phi)$ and
$b^\star_{j,\epsilon}(\phi)$ are bounded operators on $\F$ and
$$
\norm{b^\star_{j,\epsilon}(\phi)}=\norm{b_{j,\epsilon}(\phi)}=\norm{\phi}.
$$
The $b_{j,\epsilon}(\phi)$'s and the $b^\star_{j,\epsilon}(\phi)$'s 
satisfy similar anticommutation relations (see \cite{Th}).\\
The free Hamiltonian $H_{0}$ is given by
\be \label{H0}
H_{0}=\sum_{j=1}^4\sum_{\ep=+,-}\int d\xi \om_{j}(\xi)\bsjx\bjx
\ee
where
\bas
\om_{1}(\xi)&=\om_{1}(p)=\sqrt{\va{p}^{2}+m_{1}^2}\\
\om_{4}(\xi)&=\om_{4}(p)=\sqrt{\va{p}^{2}+m_{4}^2}\\
\om_{j}(\xi)&=\om_{j}(p)=\va{p}, \ j=2,3
\end{split}\end{align}
and the mass $m_{1}$ and $m_{4}$ are strictly positive. We know that 
$m_{1}<m_{4}$.\\
$H_{0}$ is essentially self-adjoint on $\F_{0}$, we still denote 
$H_{0}$ its self-adjoint extension.\\
The interaction, denoted by $H_{I}$ is given by
\basl{HI}
H_{I}=&\sum_{\ep\neq\ep'}\int d\xi_{1}d\xi_{2}d\xi_{3}d\xi_{4}
G_{\ep,\ep'}(\xi_{1},\xi_{2},\xi_{3},\xi_{4})\\
&\quad\quad b^\star_{1,\epsilon}(\xi_{1})b^\star_{2,\epsilon'}(\xi_{2})
b^\star_{3,\epsilon}(\xi_{3})b_{4,\epsilon}(\xi_{4})\\
&+\sum_{\ep\neq\ep'}\int d\xi_{1}d\xi_{2}d\xi_{3}d\xi_{4}
\overline{G_{\ep,\ep'}(\xi_{1},\xi_{2},\xi_{3},\xi_{4})}\\
&\quad\quad b^\star_{4,\epsilon}(\xi_{4})b_{3,\epsilon}(\xi_{3})
b_{2,\epsilon'}(\xi_{2})b_{1,\epsilon}(\xi_{1})
\end{split}\end{align}
where $G_{\ep,\ep'}(\xi_{1},\xi_{2},\xi_{3},\xi_{4})$ is a kernel.\\ In 
particular this interaction describes the decay of the muon $\mu$ 
into an electron and two neutrinos $\bar{\nu}_{e}$ and $\nu_{\mu}$.\\
The total Hamiltonian is then
\be \label{H}
H=H_{0}+gH_{I}
\ee
where $g\in \R$ is the coupling constant.\\
We first show that a self-adjoint operator in $\F$ is associated with 
the total Hamiltonian $H$ if the kernels $G_{\ep,\ep'}$ are in $L^2$.

\medskip

Let $\{e_{+,i},e_{-,\bar i},\ i,\bar i = 1,2,\ldots\}$ (resp.
$\{f_{+,i},f_{-,\bar i},\ i,\bar i = 1,2,\ldots\}$,
$\{g_{+,i},g_{-,\bar i},\ i,\bar i = 1,2,\ldots\}$,
$\{h_{+,i},h_{-,\bar i},\ i,\bar i = 1,2,\ldots\}$) be two basis of 
$L^2(\Sigma_{1})$ (resp. $L^2(\Sigma_{2})$, $L^2(\Sigma_{2})$, $L^2(\Sigma_{1})$).
We assume that the $e$'s, $f$'s, $g$'s and $h$'s are smooth functions 
in the Schwartz space with respect to $p$.\\
For every $Q=(q,\bar q,r,\bar r, s, \bar s, T, \bar t)\in \N^8$ we 
now consider vectors in $\F$ of the following form:
\basl{6}
\Psi^\Q=&b^\star_{1,+}(e_{+i_{1}})\ldots b^\star_{1,+}(e_{+i_{q}})
b^\star_{1,-}(e_{-\bar i_{1}})\ldots b^\star_{1,-}(e_{-\bar i_{\bar q}})\\
&b^\star_{2,+}(f_{+j_{1}})\ldots b^\star_{2,+}(f_{+j_{r}})
b^\star_{2,-}(f_{-\bar j_{1}})\ldots b^\star_{2,-}(f_{-\bar j_{\bar r}})\\
&b^\star_{3,+}(g_{+k_{1}})\ldots b^\star_{3,+}(g_{+k_{s}})
b^\star_{3,-}(g_{-\bar k_{1}})\ldots b^\star_{3,-}(g_{-\bar k_{\bar r}})\\
&b^\star_{4,+}(h_{+l_{1}})\ldots b^\star_{4,+}(h_{+l_{t}})
b^\star_{4,-}(h_{-\bar l_{1}})\ldots b^\star_{4,-}(h_{-\bar l_{\bar 
t}}) \ \Om .
\end{split}\end{align}
The indexes are ordered such that $i_{1}<\ldots<i_{q}$, $\bar 
i_{1}<\ldots<\bar i_{\bar q}$ and similarly for the indexes $j,k,l$. 
The set $\{\Psi^\Q \mid Q\in \N^8\}$ is an orthonormal basis of $\F$ 
(see \cite{Th}) and the set 
$$
\F_{\mathrm fin}=\{ \mbox{ finite linear combination of the basis 
vectors of the form \eqref{6} }\}$$
is dense in $\F$.\\
As the formal expression of $H$ shows, we have to 
deal with operators in $\F$ built from the product of creation and 
annihilation operators.\\
For $H_{\ep,\ep'}(\cdot, \cdot, \cdot)\in L^2(\Sigma_{1}\times 
\Sigma_{2}\times \Sigma_{2})$ the formal operator 
$$
\int_{\Sigma_{1}\times 
\Sigma_{2}\times \Sigma_{2}}d\xi_{1}d\xi_{2}d\xi_{3}\overline{
H_{\ep,\ep'}(\xi_{1}, \xi_{2}, \xi_{3})}b_{3,\ep}(\xi_{3})
b_{2,\ep'}(\xi_{2})b_{1,\ep}(\xi_{1})$$
is defined as a quadratic form on $\F_{\mathrm fin}\times \F_{\mathrm 
fin}$:
$$
\int_{\Sigma_{1}\times 
\Sigma_{2}\times \Sigma_{2}}d\xi_{1}d\xi_{2}d\xi_{3}<\Psi\ ,\ \overline{
H_{\ep,\ep'}(\xi_{1}, \xi_{2}, \xi_{3})}b_{3,\ep}(\xi_{3})
b_{2,\ep'}(\xi_{2})b_{1,\ep}(\xi_{1})\Phi >.$$
By mimicking the proof of Theorem X.44 in \cite{RS2}, we get an 
operator, denoted by $A_{\ep,\ep'}$, associated with the form such 
that $A_{\ep,\ep'}$ is the unique operator in $\F$ such that 
$\F_{\mathrm fin}\subset D(A_{\ep,\ep'})$ is a core for $A_{\ep,\ep'}$ and 
$$
A_{\ep,\ep'}=
\int_{\Sigma_{1}\times 
\Sigma_{2}\times \Sigma_{2}}d\xi_{1}d\xi_{2}d\xi_{3}\overline{
H_{\ep,\ep'}(\xi_{1}, \xi_{2}, \xi_{3})}b_{3,\ep}(\xi_{3})
b_{2,\ep'}(\xi_{2})b_{1,\ep}(\xi_{1})$$
as a quadratic forms on $\F_{\mathrm fin}\times \F_{\mathrm 
fin}$. Note that the formal operator
$$
\int_{\Sigma_{1}\times 
\Sigma_{2}\times \Sigma_{2}}d\xi_{1}d\xi_{2}d\xi_{3}{
H_{\ep,\ep'}(\xi_{1}, \xi_{2}, \xi_{3})}b^\star_{1,\ep}(\xi_{1})
b^\star_{2,\ep'}(\xi_{2})b^\star_{3,\ep}(\xi_{3})$$
is similarly associated with $A^\star_{\ep,\ep'}$ and we have
$$
A^\star_{\ep,\ep'}=\int_{\Sigma_{1}\times 
\Sigma_{2}\times \Sigma_{2}}d\xi_{1}d\xi_{2}d\xi_{3}{
H_{\ep,\ep'}(\xi_{1}, \xi_{2}, \xi_{3})}b^\star_{1,\ep}(\xi_{1})
b^\star_{2,\ep'}(\xi_{2})b^\star_{3,\ep}(\xi_{3})$$
as a quadratic forms on $\F_{\mathrm fin}\times \F_{\mathrm 
fin}$.\\
The proofs of the following propositions are similar to those in 
\cite{BDG}. For sake of completeness we give here complete proofs.\\
We have
\begin{proposition}\label{prop1}
Suppose that $H_{\ep,\ep'}(\cdot, \cdot, \cdot)\in L^2(\Sigma_{1}\times 
\Sigma_{2}\times \Sigma_{2})$. Then $A_{\ep,\ep')}$ and 
$A^\star_{\ep,\ep'}$ are bounded operators in $\F$ with
$$
\norm{A_{\ep,\ep'}}=\norm{A^\star_{\ep,\ep'}}\leq 
\norm{H_{\ep,\ep'}}_{L^2(\Sigma_{1}\times 
\Sigma_{2}\times \Sigma_{2})}.
$$
\end{proposition}    
\proof Let $\Psi^\Q$ be a vector of the form \eqref{6}. For simplicity 
we assume that $\{i_{1},\ldots,i_{q}\}=\{1,\ldots,q\}$, 
$\{\bar i_{1},\ldots,\bar i_{\bar q}\}=\{1,\ldots,\bar q\}$, 
etc\ldots Let us consider $A_{+,-}$, the other choices of $\ep$ and 
$\ep'$ are treated similarly. A straightforward computation shows that 
\basl{11}
A_{+,-}\Psi^\Q=& \sum_{\al=1}^q \sum_{\bet=1}^{\bar r}\sum_{\ga=1}^s
(-1)^{\al+\bet+\ga+1}(H_{+,-},e_{+\al}\otimes f_{-\bet}\otimes g_{+\ga})_{
L^2(\Sigma_{1}\times 
\Sigma_{2}\times \Sigma_{2})}\\
&\prod_{i=1\ i\neq\al}^qb^\star_{1+}(e_{+i})
\prod_{\bar i=1}^{\bar q}b^\star_{1-}(e_{-\bar i})
\prod_{j=1}^rb^\star_{2+}(f_{+j})
\prod_{\bar j=1 \ \bar k\neq \bet}^{\bar r}b^\star_{2-}(f_{-\bar j})\\
&\prod_{k=1\ k\neq\ga}^sb^\star_{3+}(g_{+k})
\prod_{\bar k=1}^{\bar s}b^\star_{3-}(g_{-\bar k})
\prod_{l=1}^tb^\star_{4+}(h_{+l})
\prod_{\bar k=1}^{\bar t}b^\star_{4-}(h_{-\bar l})\ \Om.
\end{split}\end{align}
As the right hand side of \eqref{11} is a linear combination of orthogonal vectors, 
we get
\basl{12}
\norm{A_{+,-}\Psi^\Q}^2=&\sum_{\al=1}^q \sum_{\bet=1}^{\bar r}\sum_{\ga=1}^s
\Va{(H_{+,-},e_{+\al}\otimes f_{-\bet}\otimes g_{+\ga})}^2\\
\leq & \norm{H_{+,-}}^2 \norm{\Psi^\Q}^2.
\end{split}\end{align}
Therefore, in order to prove proposition \ref{prop1}, it is enough to 
show that \eqref{12} holds for any finite linear combination of the 
$\Psi^\Q$'s. This can be done as in the proposition 3.4 of \cite{BDG}. 
We omit the details. \qed

We now investigate operators in $\F$ associated with the interaction 
$H_{I}$. Let us introduce the operators number of each particle:
\be \label{13}
N_{i}=\sum_{\ep}\int d\xi b^\star_{i\ep}(\xi)b_{i\ep}(\xi)
\quad i=1,2,3,4.\ee
Each $N_{i}$ is self-adjoint in $\F$ and $\F_{\mathrm fin}$ is a core for it.\\
For $G_{\ep,\ep'}(\cdot,\cdot, \cdot, \cdot)\in L^2(\Sigma_{1}\times 
\Sigma_{2}\times \Sigma_{2}\times \Sigma_{1})$ the formal operators 
$$
\int_{\Sigma_{1}\times 
\Sigma_{2}\times \Sigma_{2}\times \Sigma_{1}}d\xi_{1}d\xi_{2}d\xi_{3}d\xi_{4}
G_{\ep,\ep'}(\xi_{1}, \xi_{2}, \xi_{3}, \xi_{4})b^\star_{1,\ep}(\xi_{1})
b^\star_{2,\ep'}(\xi_{2})b^\star_{3,\ep}(\xi_{3})b_{4,\ep}(\xi_{4})$$
and
$$
\int_{\Sigma_{1}\times 
\Sigma_{2}\times \Sigma_{2}\times 
\Sigma_{1}}d\xi_{1}d\xi_{2}d\xi_{3}d\xi_{4}\overline{
G_{\ep,\ep'}(\xi_{1}, \xi_{2}, \xi_{3}, 
\xi_{4})}b^\star_{4,\ep}(\xi_{4})b_{3,\ep}(\xi_{3})
b_{2,\ep'}(\xi_{2})b_{1,\ep}(\xi_{1})$$
are defined as a quadratic form on $\F_{\mathrm fin}\times \F_{\mathrm 
fin}$. Again
by mimicking the proof of Theorem X.44 in \cite{RS2}, we get an 
operator, denoted by $B_{\ep,\ep'}$, associated with the form such 
that $B_{\ep,\ep'}$ is the unique operator in $\F$ such that 
$\F_{{\mathrm fin}}\subset D(A_{\ep,\ep'})$ is a core for $B_{\ep,\ep'}$ and 
$$
B_{\ep,\ep'}=
\int_{\Sigma_{1}\times 
\Sigma_{2}\times \Sigma_{2}\times \Sigma_{1}}d\xi_{1}d\xi_{2}d\xi_{3}d\xi_{4}
G_{\ep,\ep'}(\xi_{1}, \xi_{2}, \xi_{3}, \xi_{4})b^\star_{1,\ep}(\xi_{1})
b^\star_{2,\ep'}(\xi_{2})b^\star_{3,\ep}(\xi_{3})b_{4,\ep}(\xi_{4})$$
and
$$
B^\star_{\ep,\ep'}=
\int_{\Sigma_{1}\times 
\Sigma_{2}\times \Sigma_{2}\times 
\Sigma_{1}}d\xi_{1}d\xi_{2}d\xi_{3}d\xi_{4}\overline{
G_{\ep,\ep'}(\xi_{1}, \xi_{2}, \xi_{3}, 
\xi_{4})}b^\star_{4,\ep}(\xi_{4})b_{3,\ep}(\xi_{3})
b_{2,\ep'}(\xi_{2})b_{1,\ep}(\xi_{1})$$
as  quadratic forms on $\F_{\mathrm fin}\times \F_{\mathrm 
fin}$.\\
We then have
\begin{proposition}\label{prop2}
Suppose that $G_{\ep,\ep'}(\cdot,\cdot, \cdot, \cdot)\in L^2(\Sigma_{1}\times 
\Sigma_{2}\times \Sigma_{2}\times \Sigma_{1})$. Then $D(B_{\ep,\ep'}),
D(B^\star_{\ep,\ep'})\supset D(N_{4}^{1/2})$ and 
\basl{18}
\norm{B_{\ep,\ep'}\Psi}&\leq 
\norm{G_{\ep,\ep'}}_{L^2(\Sigma_{1}\times 
\Sigma_{2}\times \Sigma_{2}\times 
\Sigma_{1})}\norm{N_{4}^{1/2}\Psi},\\
\norm{B^\star_{\ep,\ep'}\Psi}&\leq 
\norm{G_{\ep,\ep'}}_{L^2(\Sigma_{1}\times 
\Sigma_{2}\times \Sigma_{2}\times 
\Sigma_{1})}\norm{N_{4}^{1/2}\Psi}.
\end{split}\end{align}
for $\Psi\in D(N_{4}^{1/2})$.
\end{proposition}    
\proof
We only investigate $B_{+,-}$. The proof for the other cases is quite 
similar. Set $Q=(q,\bar q,r,\bar r, s, \bar s, t, \bar t)$ and 
$Q'=(q+1,\bar q,r,\bar r+1, s+1, \bar s, t-1, \bar t)$.
Let $\Pq$ and $\Pqp$ be two vectors in $\F_{{\mathrm fin}}\cap \F^\Q$ 
and $\F_{{\mathrm fin}}\cap \F^{(Q')}$ respectively.\\
We have
\bas
&(\Pqp, B_{+,-}\Pq)=
\int_{\Sigma_{1}\times 
\Sigma_{2}\times \Sigma_{2}\times 
\Sigma_{1}}d\xi_{1}d\xi_{2}d\xi_{3}d\xi_{4}\\
&\left(\overline{
G_{+,-}(\xi_{1}, \xi_{2}, \xi_{3}, 
\xi_{4})}b_{3,+}(\xi_{3})
b_{2,-}(\xi_{2})b_{1,+}(\xi_{1})\Pqp\ ,\ 
b_{4,+}(\xi_{4})\Pq\right)
\end{split}\end{align}
and by the Fubini theorem, we get
\bas
&\Va{(\Pqp, B_{+,-}\Pq)}^2=
\Big|\int_{\Sigma_{1}}d\xi_{4}\Big( b_{4,+}(\xi_{4})\Pq, \\
 &\int_{\Sigma_{1}\times 
\Sigma_{2}\times \Sigma_{2}}\overline{
G_{+,-}(\xi_{1}, \xi_{2}, \xi_{3}, 
\xi_{4})}b_{3,+}(\xi_{3})
b_{2,-}(\xi_{2})b_{1,+}(\xi_{1})\Pqp \Big)\Big|^2.
\end{split}\end{align}
By the Cauchy-Schwarz inequality and proposition \ref{prop1}, we 
obtain
\bas
&\Va{(\Pqp, B_{+,-}\Pq)}^2\leq \\
&\left(\int_{\Sigma_{1}}d\xi_{4}\norm{ b_{4,+}(\xi_{4})\Pq}
\big(\int_{\Sigma_{1}\times 
\Sigma_{2}\times \Sigma_{2}}d\xi_{1}d\xi_{2}d\xi_{3}\va{
G_{+,-}(\xi_{1}, \xi_{2}, \xi_{3}, 
\xi_{4})}^2\big)^{1/2} \right)^2 \norm{\Pqp}^2.
\end{split}\end{align}
Applying again the Cauchy-Schwarz inequality and by the definition of 
$b_{4+}(\xi_{4})$ we finally get
$$
\Va{(\Pqp, B_{+,-}\Pq)}^2\leq 
t\norm{G_{+,-}}^{2}\norm{\Pq}^2\norm{\Pqp}^2=
\norm{G_{+,-}}^{2}\norm{N_{4}^{1/2}\Pq}^2\norm{\Pqp}^2.$$
Since $B_{+,-}\Pq\in \F^{(Q')}$ we deduce
$$
\Va{(\Phi, B_{+,-}\Pq)}^2\leq 
\norm{G_{+,-}}^{2}\norm{N_{4}^{1/2}\Pq}^2\norm{\Phi}^2$$
for every $\Phi \in \F_{{\mathrm fin}}$. Now, since $\Phi \in \F_{{\mathrm 
fin}}$ is dense in $\F$, the last inequality  still holds for every 
$\Phi\in \F$ and every $Q\in \N^8$. Therefore we have
$$
\norm{B_{+,-}\Pq}^2\leq \norm{G_{+,-}}^{2}\norm{N_{4}^{1/2}\Pq}^2
$$
which yields
\be \label{26}
\norm{B_{+,-}\Psi}^2\leq \norm{G_{+,-}}^{2}\norm{N_{4}^{1/2}\Psi}^2
\ee
for every $\Psi\in \F_{{\mathrm fin}}$. Since $\F_{{\mathrm fin}}$ is 
a core for $N_{4}^{1/2}$ and $B_{+,-}$ is closable (see Theorem X.44 
in \cite{RS2}) we have $D(N_{4}^{1/2})\subset D(B_{+,-})$ and the 
inequality \eqref{26} is still true for every $\Psi \in 
D(N_{4}^{1/2})$. \qed

Set 
\bas
V_{2}^{\ep\ep'}=&\int_{\Sigma_{1}\times 
\Sigma_{2}\times 
\Sigma_{1}}d\xi_{1}d\xi_{3}d\xi_{4}G_{\ep\ep'}^2(\xi_{1},\xi_{3},\xi_{4})\\
&b_{1\ep}^\star(\xi_{1})b_{3\ep'}^\star(\xi_{3})
b_{4\ep}(\xi_{4}),\\
V_{3}^{\ep\ep'}=&\int_{\Sigma_{1}\times 
\Sigma_{2}\times 
\Sigma_{1}}d\xi_{1}d\xi_{2}d\xi_{4}G_{\ep\ep'}^3(\xi_{1},\xi_{2},\xi_{4})\\
&b_{1\ep}^\star(\xi_{1})b_{2\ep'}^\star(\xi_{2})
b_{4\ep}(\xi_{4}),
\end{split}\end{align}
where $G_{\ep\ep'}^j \in L^2 (\Sigma_{1}\times 
\Sigma_{2}\times 
\Sigma_{1})$, $j=2,3$.
$V_{j}^{\ep\ep'}$, $j=2,3$, are defined as quadratic forms on 
$\F_{{\mathrm fin}}\times\F_{{\mathrm fin}}$. As above we then have
\begin{proposition}\label{prop2bis}
    Suppose that $G^j_{\ep\ep'}\in L^2(\Sigma_{1}\times 
    \Sigma_{2}\times \Sigma_{1})$, $j=2,3$. Then $D(V^{\ep\ep'}_{j}),
    D(V_{j}^{\ep\ep'\star})\supset D(N_{4}^{1/2})$ and 
    \basl{18bis}
    \norm{V^{\ep\ep'}_{j}\Psi}&\leq 
    \norm{G^j_{\ep\ep'}}_{L^2(\Sigma_{1}\times 
    \Sigma_{2}\times 
    \Sigma_{1})}\norm{N_{4}^{1/2}\Psi},\\
    \norm{V^{\ep\ep'\star}_{j}\Psi}&\leq 
    \norm{G^j_{\ep\ep'}}_{L^2(\Sigma_{1}\times 
    \Sigma_{2}\times 
    \Sigma_{1})}\norm{N_{4}^{1/2}\Psi}.
    \end{split}\end{align}
    for $\Psi\in D(N_{4}^{1/2})$ and $j=2,3$.
    \end{proposition} 
The proof of proposition \ref{prop2bis} is exactly the same as the one 
of proposition \ref{prop2}.

\medskip

The following theorem shows that the formal total Hamiltonian is 
associated with a self-adjoint operator in $\F$, still denoted by 
$H$, if the interaction kernels are in $L^{2}$.
\begin{theorem}\label{thm3}
Suppose that $G_{\ep\ep'}(\cdot,\cdot,\cdot,\cdot)\in L^2(
{\Sigma_{1}\times 
\Sigma_{2}\times\Sigma_{2}\times \Sigma_{1}})$ for $\ep\neq \ep'$. 
Then $H=H_{0}+gH_{I}$ is a self-adjoint operator in $\F$ for every 
$g\in \R$ with domain $D(H_{0})$.
\end{theorem}
\proof  Recall that $H_{0}$ with domain $\Ff$  is essentially 
self-adjoint. By proposition \ref{prop2} we have, for every $\Psi 
\in \Ff$,
$$ \norm{H_{I}\Psi}\leq 2\left( 
\sum_{\ep\neq\ep'}\norm{G_{\ep\ep'}}_{L^2}\right)
\norm{N_{4}^{1/2}\Psi}$$
and we get for every $\ep >0$,
$$ \norm{H_{I}\Psi}\leq 2\left( 
\sum_{\ep\neq\ep'}\norm{G_{\ep\ep'}}_{L^2}\right)\left( 
\sqrt{\ep/2}\norm{N_{4}\Psi}+ 
\frac{1}{\sqrt{2\ep}}\norm{\Psi}\right).$$
Furthermore, since $\om_{4}(p)\geq m_{4}$, we have
$$\norm{N_{4}\Psi}\leq \frac{1}{m_{4}}\norm{H_{0}\Psi}.$$
Thus
$$\norm{H_{I}\Psi}\leq 2\left( 
\sum_{\ep\neq\ep'}\norm{G_{\ep\ep'}}_{L^2}\right)\left( 
\frac{1}{m_{4}}\sqrt{\ep/2}\norm{H_{0}\Psi}+ 
\frac{1}{\sqrt{2\ep}}\norm{\Psi}\right)$$
which means that $H_{I}$ is relatively bounded with respect to 
$H_{0}$ with zero relative bound and the theorem follows from the 
Kato-Rellich theorem. \qed

\section{The results}
Our main result states that $H$ has a ground state for $g$ 
sufficiently small. We have
\begin{theorem}\label{thm4}
Suppose that for $\ep\neq\ep'$, 
$G_{\ep\ep'}(\cdot,\cdot,\cdot,\cdot)\in 
L^2(\Sigma_{1},\Sigma_{2},\Sigma_{2},\Sigma_{1})$ and
\basl{H1}
\sum_{i=2}^3\int_{\overline{B(0,1)}}\frac{\Va{G_{\ep\ep'}
(\xi_{1},\xi_{2},\xi_{3},\xi_{4})}^2}{\va{p_{i}}^2}
d\xi_{1}d\xi_{2}d\xi_{3}d\xi_{4}<\infty
\end{split}\end{align}
where $\xi_{j}=(p_{j},s_{j})$, $p_{j}\in \R^3$, $j=1,2,3,4$ and where 
$\overline{B(0,1)}=\{ (p_{1},p_{2},p_{3},p_{4})\in \R^{12}\mid 
\sum_{j=1}^4\va{p_{j}}^2\leq 1\}$.\\
Then there exists $g_{0}>0$ such that $H$ has an unique ground state 
for $\va{g}\leq g_{0}$. Furthermore 
$\sigma(H)=\sigma_{\mathrm{ac}}(H)=[\inf\sigma(H),+\infty)$.
\end{theorem}
Notice that Theorem \ref{thm4} is true for sharp cutoffs, i.e., when
$G_{\ep\ep'}=\chi_{\Lambda}$, $\Lambda >0$, with
\bas \chi_{\Lambda}(p_{1},p_{2},p_{3},p_{4})&=1 \mbox{ if } \va{p_{j}}\leq 
\Lambda,\ j=1,2,3,4\\
&=0 \mbox{ otherwise.}
\end{split}\end{align}
This means that the ground state exists without infrared 
regularization even if particles with zero mass are involved.\\
The statement concerning the absolutely continuous spectrum of $H$ 
follows easily from the existence of asymptotic Fock representations 
of the ACR. Precisely, for $f\in L^2(\R^3)$ we define the operators
$$
b_{j\ep,t}^\flat (f)=e^{itH}e^{-itH_{0}}b_{j\ep}^\flat (f)
e^{itH_{0}}e^{-itH},\quad j=1,2,3,4\ ,\ep=\pm.$$
Then for $f\in C_{0}^\infty (\R^3)$ and $\psi\in \F$ the strong 
limits of $b_{j\ep,t}^\flat (f)$ exist:
$$
\lim_{t\to \pm \infty}b_{j\ep,t}^\flat (f)\psi= b_{j\ep,\pm}^\flat 
(f)\psi.$$
The $b_{j\ep,\pm}^\flat (f)$'s satisfy the ACR and if $\phi$ is the 
ground state of $H$, we have, for $f\in C_{0}^\infty (\R^3)$,
$$
b_{j\ep,\pm}^\flat (f) \phi =0.$$
The fact that 
$\sigma(H)=\sigma_{\mathrm{ac}}(H)=[\inf\sigma(H),+\infty)$ follows 
by mimicking \cite{H04}.\\
Now the next theorem concerns the absolutely continuous 
spectrum of $H$. We define $S$ as the set of threshold of $H_{0}$:
\be \label{S}
S=\{km_{1}+lm_{4}\mid k,l\in \N\}.
\ee
\begin{theorem}\label{thm5}
    Suppose that for $\ep\neq\ep'$, 
    $G_{\ep\ep'}(\cdot,\cdot,\cdot,\cdot)\in 
    L^2(\Sigma_{1}\times\Sigma_{2}\times\Sigma_{2}\times\Sigma_{1})$ satisfy 
    \eqref{H1} and that for $i=1,2,3,4$,
    $p_{i}\cdot \nabla_{p_{i}}G_{\ep\ep'}$ and 
    $p_{i}^2\Delta_{p_{i}}G_{\ep\ep'}$ are all in 
    $L^2(\Sigma_{1}\times\Sigma_{2}\times\Sigma_{2}\times\Sigma_{1})$.
    Then there exists a constant $C>0$ such 
    that, for $g$ sufficiently small,  
    the spectrum of $H$ in $\R\setminus (S+[-C\sqrt g,C\sqrt g])$ 
    is absolutely continuous.
\end{theorem}

\section{Proof of theorem \ref{thm4}}

Let $\his$ be the operator obtained from \eqref{HI} by substituting 
$$
G^\sigma_{\ep\ep'}(\xi_{1},\xi_{2},\xi_{3},\xi_{4})=
1_{\{(p_{1},p_{2},p_{3},p_{4})\mid \va{p_{2}\geq 
\sigma,\, \va{p_{3}}\geq 
\sigma}\}}G_{\ep\ep'}(\xi_{1},\xi_{2},\xi_{3},\xi_{4})$$
for $G_{\ep\ep'}$ where $\sigma$ is a strictly positive parameter.
We then define 
$$
H_{\sigma}=H_{0}+g\his .$$
$H_{\sigma}$ is a self adjoint operator in $\F$ with domain 
$D(H_{\sigma})=D(H_{0})$ for any $g\in \R$ and any $\sigma >0$.\\
Set 
\be \label{h10}
H^1_{0}=\sum_{\ep}\int \om_{1}(\xi)b^\star_{1\ep}(\xi)b_{1\ep}(\xi) 
d\xi + \sum_{\ep}\int \om_{4}(\xi)b^\star_{4\ep}(\xi)b_{4\ep}(\xi) 
d\xi. \ee
We consider $H^1_{0}$ as a self-adjoint operator in the Fock space 
$\F_{1}$ associated with the particles and antiparticles 1 and 4. We 
then have $\sigma(H^1_{0})=\{0\}\cup [m_{1}, +\infty)$ because 
$m_{1}<m_{4}$.\\
For $0<\la <m_{1}$ let $P(\la)$ be the spectral projection of 
$H^1_{0}$ in $\F_{1}$ corresponding to $(-\infty, \la]$ and let 
$P_{\Om_{\mathrm neut}}$ be the orthogonal projection on the vacuum 
state of the neutrinos and antineutrinos 2 and 3. We consider $\pon$ as 
a projection in the Fock space $\F_{2}$ associated with the neutrinos 
and antineutrinos 2 and 3. Note that $\F\equiv \F_{1}\otimes \F_{2}$. 
As in \cite{BDG} and \cite{BFS98b} theorem \ref{thm4} is the consequence of the 
following theorem:
\begin{theorem}\label{thm6}
There exists $g_{0}>0$ such that for every $g$ satisfying $\vag\leq 
g_{0}$ the following properties hold:
\begin{itemize}
\item[(i)] For every $\psi \in D(H_{0})$ we have $H_{\sigma}\psi \to 
H\psi$ as $\sigma \to 0$.
\item[(ii)] For every $\sigma\in (0,1]$, $H_{\sigma}$ has a 
normalized ground state $\phi_{\sigma}$.
\item[(iii)] We have for every $\sigma \in (0,1]$
$$
(\phi_{\sigma},P(\la)\otimes \pon \phi_{\sigma})\geq 1-\delta_{g}(\la)
$$
where $\delta_{g}(\la)$ tends to zero when $g$ tends to zero and 
$0\leq\delta_{g}(\la)< 1$ for $\vag \leq g_{0}$.
\end{itemize}    
\end{theorem}
\proof  We first estimate $E_{\sigma}= \inf\sigma(H_{\sigma})$, ${\sigma \in 
(0,1]}$. One proves that $E_{\sigma}\leq 0$ as in lemma 4.3 of 
\cite{BDG}. \\
Recall that there exist a constant $C>0$ such that for every $\eta >0$ 
and
for every $\sigma\in (0,1]$
\be \label{eta}
\norm{\his\psi}\leq C(\sqrt \eta \norm{H_{0}\psi}+\frac{1}{\sqrt 
\eta}\norm{\psi}),\quad \psi \in D(H_{0}).
\ee
Therefore it follows from the Kato-Rellich theorem that
\be\label{37}
\va{E_{\sigma}}\leq \frac{\vag C}{\sqrt \eta -\vag \eta C}
\ee
when $\vag \sqrt \eta C<1$.\\
(i) follows from the following inequality and from the Lebesgue's 
theorem:
$$\norm{(H-H_{\sigma})\psi}\leq 2 C\vag (\sum_{\ep\neq 
\ep'}\norm{G_{\ep\ep'}-G^\sigma_{\ep\ep'}}_{L^2})
(\sqrt \eta \norm{H_{0}\psi}+\frac{1}{\sqrt 
\eta}\norm{\psi}).$$
(ii) is proved as in \cite{BFS98b} or in \cite{BDG} (theorem 4.10). 
We omit the details. Thus we have 
$H_{\sigma}\phi_{\sigma}=E_{\sigma}\phi_{\sigma}$ with 
$\norm{\phi_{\sigma}}=1$.\\
Writing 
$H_{0}\phi_{\sigma}=H_{\sigma}\phi_{\sigma}-g\his\phi_{\sigma}$ we get 
using \eqref{eta} and \eqref{37}
\basl{39}
\norm{H_{0}\phi_{\sigma}}&\leq (\va{E_{\sigma}}+\vag \frac{C}{\sqrt 
\eta})(1 -\sqrt\eta \vag C)^{-1}\\
&\leq \vag \frac{C}{\sqrt \eta})(1 
-\sqrt\eta \vag C)^{-2} (2 -\vag  \sqrt\eta C)
\end{split}\end{align}
for every $\sigma\in(0,1]$ and for $\sqrt\eta \vag C<1$.\\
It remains to prove (iii). Note that (iii) is equivalent to
\be \label{39'}
((P(\la)^\perp\otimes \pon 
+1\otimes\pon^\perp)\phi_{\sigma},\phi_{\sigma})\leq \delta_{g}(\la)
\ee
for every $\sigma\in(0,1]$.\\
Note that 
\bas
0=&(P(\la)^\perp\otimes \pon )(H_{\sigma}-E_{\sigma})\phi_{\sigma}\\
=&P(\la)^\perp (H^1_{0}\otimes 1-E_{\sigma})\otimes \pon \phi_{\sigma}
+g(P(\la)^\perp\otimes\pon)\his \phi_{\sigma}.
\end{split}\end{align}
Remarking that $P(\la)^\perp H^1_{0}\geq m_{1}P(\la)^\perp$ and using 
$E_{\sigma}\leq 0$,
we get 
$$
(P(\la)^\perp\otimes \pon 
\phi_{\sigma},\phi_{\sigma})\leq -\frac{\vag}{m_{1}}
(P(\la)^\perp\otimes \pon \his
\phi_{\sigma},\phi_{\sigma}).$$
Furthermore it follows from \eqref{eta} that there exists a constant 
$C>0$ such that
$$\Va{(P(\la)^\perp\otimes \pon \his
\phi_{\sigma},\phi_{\sigma})}\leq C$$
and thus
\be\label{43}
(P(\la)^\perp\otimes \pon 
\phi_{\sigma},\phi_{\sigma})\leq C \frac{\vag}{m_{1}}
\ee
On the other hand one easily verifies that there exists a constant 
$C>0$ such that 
\be \label{44}
\norm{\pon^\perp \phi_{\sigma}}\leq C(\norm{N_{2}^{1/2}\phi_{\sigma}}
+\norm{N_{3}^{1/2}\phi_{\sigma}})
\ee
for every $\sigma\in(0,1]$ where we recall that $N_{j}=\sum_{\ep}\int 
b_{j\ep}^\star(\xi)\bjx d\xi$.\\
The proof of (iii) then follows from \eqref{39'}, \eqref{43}, \eqref{44} 
and the following lemma
\begin{lemma}\label{lem6}
There exists a constant $C>0$ such that 
\be\label{45}
\norm{N_{j}^{1/2}\phi_{\sigma}}^2\leq g^2C\left(\sum_{\ep\neq\ep'}\int
\frac{\va{G_{\ep\ep'}(\xi_{1},\xi_{2},\xi_{3},\xi_{4})}^2}{\va{p_{j}}^2}
d\xi_{1}d\xi_{2}d\xi_{3}d\xi_{4}\right)\norm{H_{0}\phi_{\sigma}}^2\ee
for $j=2,3$ and for every $\sigma\in (0,1]$.
\end{lemma}
\proof Recall that,
\be \label{44bis}
\{b_{2\ep}(\xi),b_{3\ep}^\flat(\xi')\}=\{b_{2\ep}(\xi),b_{3\ep'}^\flat(\xi')\}=0
\ee
according to our convention. It follows from the CAR and \eqref{44bis} that we have the following pull-through formula:
$$
0=(H_{\sigma}-E_{\sigma}+\om_{j}(\xi))\bjx\phi_{\sigma} + 
gV_{j}^{\ep\ep'\sigma}(\xi)\phi_{\sigma}, \quad j=2,3$$
where for $\ep\neq\ep'$
\bas
 V_{2}^{\ep\ep'\sigma}(\xi)&=\int d\xi_{1}d\xi_{3}d\xi_{4}
G^\sigma_{\ep'\ep}(\xi_{1},\xi,\xi_{3},\xi_{4})
b_{1\ep'}^\star(\xi_{1})b_{3\ep'}^\star(\xi_{3})b_{4\ep'}(\xi_{4})\\
V_{3}^{\ep\ep'\sigma}(\xi)&=\int d\xi_{1}d\xi_{2}d\xi_{4}
G^\sigma_{\ep\ep'}(\xi_{1},\xi_{2},\xi,\xi_{4})
b_{1\ep}^\star(\xi_{1})b_{2\ep'}^\star(\xi_{2})b_{4\ep}(\xi_{4}).
\end{split}\end{align}
We have
$$
\bjx\phi_{\sigma}=-g(H_{\sigma}-E_{\sigma}+\om_{j}(\xi))^{-1}
V_{j}^{\ep\ep'\sigma}(\xi)\phi_{\sigma}.$$
By proposition \ref{prop2bis} we get
\be\label{51}
\norm{b_{2\ep}(\xi)\phi_{\sigma}}^2\leq \frac{g^2}{m_{4}^2 \va{p_{2}}^2}
\left(\int
\va{G_{\ep'\ep}(\xi_{1},\xi,\xi_{3},\xi_{4})}^2
d\xi_{1}d\xi_{3}d\xi_{4}\right)\norm{H_{0}\phi_{\sigma}}^2
\ee
and
\be\label{52}
\norm{b_{3\ep}(\xi)\phi_{\sigma}}^2\leq \frac{g^2}{m_{4}^2 \va{p_{3}}^2}
\left(\int
\va{G_{\ep\ep'}(\xi_{1},\xi_{2},\xi,\xi_{4})}^2
d\xi_{1}d\xi_{2}d\xi_{4}\right)\norm{H_{0}\phi_{\sigma}}^2.
\ee
Note that
\be\label{53}
\sum_{\ep}\int \norm{\bjx 
\phi_{\sigma}}^2d\xi=\norm{N_{j}^{1/2}\phi_{\sigma}}^2\quad j=2,3.
\ee
The lemma then follows from \eqref{51}, \eqref{52} and \eqref{53} and theorem 
\ref{thm5} is proved. Note that the uniqueness (up to a phase) of the 
ground state follows as in \cite{AGG06b} and \cite{H04}. Thus theorem 
\ref{thm4} is proved.\qed

Let us remark that the proof of lemma \ref{lem6} is rather formal 
but, by mimicking \cite{H04}, one easily gets a rigorous proof. We 
omit the details.

\section{Proof of theorem \ref{thm5}}
In order to prove the absence of continuous singular spectrum away 
from the thresholds of $H_{0}$, we use the Mourre's method originates 
from \cite{Mou}. Actually this method has been applied successfuly to 
QED models (see for instance  \cite{BFS98b, BFSS99, GGM04a, GGM04b, 
Am04}). \\
To this end, we estimate from below the commutator of $H$ with an 
anti-selfadjoint operator $A=-A^\star$. Our choice for $A$ is the sum 
of the second quantization of dilatation generator on each particle 
and antiparticle space. Namely, denoting $a_{j}=\left( 
p_{j}\cdot\nabla_{p_{j}}+\nabla_{p_{j}}\cdot p_{j}\right)$, the 
generator of dilatation in the particle $j$ acting on $L^2(\R^3)$, we set
\be \label{A}
A=\sum_{\ep=\pm}\sum_{j=1}^4 d\Gamma_{j\ep}(a_{j})
\ee
where giving an operator $a$ on $L^2(\R^3)$, the operator 
$d\Gamma_{j\ep}(a): \F\to\F$ is defined by
\be
d\Gamma_{j\ep}(a)=\int d\xi b^\star_{j\ep}(\xi)\, a\, b_{j\ep}(\xi).
\ee
Note that $iA$ is essentially self-adjoint on $\Ff$.
It remains to compute $[A,H]$. 
We begin with the remark that the second quantization respects 
commutators, i.e., for given operators $a,\ a'$ on the one particle 
space $L^{2}(\R^3)$ and given $f\in L^2(\R^3)$ such that $af$ and 
$a^\star f$ belong to $L^2(\R^3)$, we have for 
$j=1,2,3,4$ and $\ep=\pm$:
\basl{aa'}
[d\Gamma_{j\ep}(a),d\Gamma_{j\ep}(a') ]\psi&=d\Gamma_{j\ep}([a,a')]\psi\\
[d\Gamma_{j\ep}(a),b^\star_{j\ep}(f)]\psi&=b^\star_{j\ep}(af)\psi\\
[d\Gamma_{j\ep}(a),b_{j\ep}(f)]\psi&=-b_{j\ep}(a^\star f)\psi,
\end{split}\end{align}
and also for $i,j=1,2,3,4$ and $\ep,\ep'=\pm$ with 
$(j,\ep)\neq(i,\ep')$:
\basl{aa''}
[d\Gamma_{j\ep}(a),d\Gamma_{i\ep'}(a') ]\psi&=0\\
[d\Gamma_{j\ep}(a),b^\star_{i\ep'}(f)]\psi&=0\\
[d\Gamma_{j\ep}(a),b_{i\ep'}(f)]\psi&=0
\end{split}\end{align}
for every $\psi \in \F_{\mathrm{fin}}$.\\
Recall that 
$$
H_{0}=\sum_{\ep=\pm}\sum_{j=1}^4 d\Gamma_{j\ep}(\om_{j})$$
and a straightforward calculus leads to
\be \label{H0A}
[A,H_{0}]\psi=\Big(\sum_{\ep=\pm}d\Gamma_{1\ep}\left(\frac{p^2}{\sqrt{p^2+m_{1}^2}}\right)
+d\Gamma_{2\ep}(\va{p})+d\Gamma_{3\ep}(\va{p})
+d\Gamma_{4\ep}\left(\frac{p^2}{\sqrt{p^2+m_{4}^2}}\right)\Big)\psi
\ee
for $\psi \in \F_{\mathrm{fin}}$.\\
Let us remark that $[A,H_{0}]$ is relatively bounded with respect to 
$H_{0}$.
\begin{proposition}\label{prop5.1}
Let $\Delta$ be a closed subset of $\R$ such that $\Delta \cap 
S=\emptyset$ and set $\beta =\mbox{dist}(\Delta,S)>0$. Then
$$
E_{\Delta}(H_{0})[A,H_{0}]E_{\Delta}(H_{0})\geq \beta E_{\Delta}(H_{0})
$$
where $E_{\Delta}(H_{0})$ denotes the spectral projection of $H_{0}$ 
for the interval 
$\Delta$.
\end{proposition}
\proof Using \eqref{H0A}, we have for a given state $\Pq\in \Fq$ such 
that  $E_{\Delta}(H_{0})\Pq=\Pq$,
\basl{H0Ap}
[A,H_{0}]\Pq (\Xi_{q},\ldots,\Xi_{\bar t})&=
\Big(\sum_{j=1}^q \frac{p_{1j}^2}{\sqrt{p_{1j}^2+m_{1}^2}}+
\sum_{j=1}^{\bar q} \frac{\bar p_{1j}^2}{\sqrt{\bar p_{1j}^2+m_{1}^2}}\\
&+\sum_{j=1}^r \va{p_{2j}}+
\sum_{j=1}^{\bar r} \va{\bar p_{2j}}\\
&+\sum_{j=1}^s \va{p_{3j}}+
\sum_{j=1}^{\bar s} \va{\bar p_{3j}}\\
&+\sum_{j=1}^t \frac{p_{4j}^2}{\sqrt{p_{4j}^2+m_{4}^2}}+
\sum_{j=1}^{\bar t} \frac{\bar p_{4j}^2}{\sqrt{\bar 
p_{4j}^2+m_{4}^2}}\Big)
\Pq (\Xi_{q},\ldots,\Xi_{\bar t}).
\end{split}\end{align}
The free energy of such state $\Pq$ is given by
\basl{E0}
H_{0}\Pq (\Xi_{q}&,\ldots,\Xi_{\bar t})=\Big(
\sum_{j=1}^q {\sqrt{p_{1j}^2+m_{1}^2}}+
\sum_{j=1}^{\bar q} {\sqrt{\bar p_{1j}^2+m_{1}^2}}\\
&+\sum_{j=1}^r \va{p_{2j}}+
\sum_{j=1}^{\bar r} \va{\bar p_{2j}}\\
&+\sum_{j=1}^s \va{p_{3j}}+
\sum_{j=1}^{\bar s} \va{\bar p_{3j}}\\
&+\sum_{j=1}^t {\sqrt{p_{4j}^2+m_{4}^2}}+
\sum_{j=1}^{\bar t} {\sqrt{\bar p_{4j}^2+m_{4}^2}}\Big)
\Pq (\Xi_{q},\ldots,\Xi_{\bar t})
\end{split}\end{align}
with
\basl{5.7bis}
\sum_{j=1}^q {\sqrt{p_{1j}^2+m_{1}^2}}+&
\sum_{j=1}^{\bar q} {\sqrt{\bar p_{1j}^2+m_{1}^2}}
+\sum_{j=1}^r \va{p_{2j}}+
\sum_{j=1}^{\bar r} \va{\bar p_{2j}}\\
+\sum_{j=1}^s \va{p_{3j}}+
\sum_{j=1}^{\bar s} \va{\bar p_{3j}}
&+\sum_{j=1}^t {\sqrt{p_{4j}^2+m_{4}^2}}+
\sum_{j=1}^{\bar t} {\sqrt{\bar p_{4j}^2+m_{4}^2}} \ \in \Delta.
\end{split}\end{align}
We decompose $H_{0}\Pq $ as follows
\basl{5.8}
H_{0}\Pq (\Xi_{q},&\ldots,\Xi_{\bar t})=\Big[(q+\bar q)m_{1}+ (t+\bar 
t)m_{4}\\
&+\sum_{j=1}^q (\sqrt{p_{1j}^2+m_{1}^2}-m_{1})+
\sum_{j=1}^{\bar q} (\sqrt{\bar p_{1j}^2+m_{1}^2}-m_{1})\\
&+\sum_{j=1}^r \va{p_{2j}}+
\sum_{j=1}^{\bar r} \va{\bar p_{2j}}\\
&+\sum_{j=1}^s \va{p_{3j}}+
\sum_{j=1}^{\bar s} \va{\bar p_{3j}}\\
&+\sum_{j=1}^t (\sqrt{p_{4j}^2+m_{4}^2}-m_{4})+
\sum_{j=1}^{\bar t} (\sqrt{\bar p_{4j}^2+m_{4}^2}-m_{4})\Big] 
\Pq (\Xi_{q},\ldots,\Xi_{\bar t}).
\end{split}\end{align}
By \eqref{5.8} we get according to the definition 
of $\beta$
\bas
&\sum_{j=1}^q (\sqrt{p_{1j}^2+m_{1}^2}-m_{1})+
\sum_{j=1}^{\bar q} (\sqrt{\bar p_{1j}^2+m_{1}^2}-m_{1})\\
+&\sum_{j=1}^r \va{p_{2j}}+
\sum_{j=1}^{\bar r} \va{\bar p_{2j}}
+\sum_{j=1}^s \va{p_{3j}}+
\sum_{j=1}^{\bar s} \va{\bar p_{3j}}\\
+&\sum_{j=1}^t (\sqrt{p_{4j}^2+m_{4}^2}-m_{4})+
\sum_{j=1}^{\bar t} (\sqrt{\bar p_{4j}^2+m_{4}^2}-m_{4})\geq \beta
\end{split}\end{align}
for $(p_{1},p_{2},p_{3},p_{4})$ satisfying \eqref{5.7bis}.\\
Therefore using
\bas
\frac{p^2}{\sqrt{p^2+m^2}}&= (\sqrt{p^2+m^2}-m)\frac{\sqrt{p^2+m^2}+m}
{\sqrt{p^2+m^2}}\\
&\geq \sqrt{p^2+m^2}-m
\end{split}\end{align}
we  conclude the proof.
\qed

\medskip

We now estimate the commutator $[A,H_{I}]$. By \eqref{HI}, \eqref{aa'}, 
\eqref{aa''} and since $a_{j}^\star =-a_{j}$ we have 
\basl{HIA}
[A,H_{I}]\psi=&\Big(\sum_{\ep\neq\ep'}\sum_{j=1}^4\int d\xi_{1}d\xi_{2}d\xi_{3}d\xi_{4}
(a_{j}G_{\ep\ep'})(\xi_{1},\xi_{2},\xi_{3},\xi_{4})\\
&\quad\quad b^\star_{1,\epsilon}(\xi_{1})b^\star_{2,\epsilon'}(\xi_{2})
b^\star_{3,\epsilon}(\xi_{3})b_{4,\epsilon}(\xi_{4})\\
+&\sum_{\ep\neq\ep'}\sum_{j=1}^4\int d\xi_{1}d\xi_{2}d\xi_{3}d\xi_{4}
\overline{(a_{j} G_{\ep\ep'})(\xi_{1},\xi_{2},\xi_{3},\xi_{4})}\\
&\quad\quad b^\star_{4,\epsilon}(\xi_{4})b_{3,\epsilon}(\xi_{3})
b_{2,\epsilon'}(\xi_{2})b_{1,\epsilon}(\xi_{1})\Big)\psi
\end{split}\end{align}
for $\psi \in \F_{\mathrm{fin}}$.
Therefore, if we assume that $a_{j}G_{\ep\ep'}\in L^2$ for each 
$j=1,2,3,4$ and for each $\ep\neq \ep'$, we deduce as 
in the proof of theorem \ref{thm3} that $[A,H_{I}]$ is $H_{0}$ 
relatively bounded and in particular there exist $c>0$ such that
\be \label{ahi}
E_{\Delta}(H_{0})[A,H_{I}]E_{\Delta}(H_{0})\geq -c E_{\Delta}(H_{0}).
\ee
We deduce
\begin{proposition}\label{prop5.2}
Assume that $a_{j}G_{\ep\ep'}\in L^2$ for each 
$j=1,2,3,4$ and for each $\ep\neq \ep'$. There exists $c>0$ such that
if $\Delta$ is a closed interval of $\R$ verifying $\Delta \cap 
S=\emptyset$ then
$$
E_{\Delta}(H)[A,H]E_{\Delta}(H)\geq (\frac \beta 2 -\frac{cg}{\beta}) E_{\Delta}(H)
$$
where $E_{\Delta}(H)$ denotes the spectral projection of $H$ 
for the interval 
$\Delta$ and $\beta =\mbox{dist}(\Delta,S)>0$ is 
suficiently small.
\end{proposition}
\proof 
Let $\Delta'$ be the closed interval such that $\Delta=\Delta' 
+[-\beta/2, \beta/2]$ and assume that $0<\beta <1$. Using the Helffer-Sj\"ostrand Functional Calculus 
(see for instance \cite{DiSj99}), we find that 
$$
\norm{E_{\Delta}(H)(1-E_{\Delta'}(H_{0}))}\leq \frac{c_{1}g}{\beta}
$$
for some  constant $c_{1}>0$ independent of 
$\Delta$, $g$ and $\beta$.\\
Therefore, using that $[A,H]$ is $H$ bounded (see the proof of theorem 
\ref{thm5} just  below),
\be\label{dd'}
E_{\Delta}(H)[A,H]E_{\Delta}(H)\geq
E_{\Delta}(H)E_{\Delta'}(H_{0})[A,H]E_{\Delta'}(H_{0})E_{\Delta}(H)-
c_{2}\frac g \beta E_{\Delta}(H)
\end{equation}
for some constant $c_{2}>0$.\\
On the other hand, from proposition \ref{prop5.1} and \eqref{ahi}, we have
\be \label{ahi2}
E_{\Delta'}(H_{0})[A,H]E_{\Delta'}(H_{0})\geq (\frac \beta 2 -c_{3}g) 
E_{\Delta'}(H_{0})
\ee
for some constant $c_{3}>0$.\\
Inserting \eqref{ahi2} in \eqref{dd'} we get
\begin{align*}\begin{split}
E_{\Delta}(H)[A,H]E_{\Delta}(H)
&\geq (\frac \beta 2 -c_{3}g)E_{\Delta}(H) E_{\Delta'}(H_{0})E_{\Delta}(H)-
c_{2}\frac g \beta E_{\Delta}(H) \\
&\geq (\frac \beta 2 -c_{3}g)(1- \frac{c_{1}g}{\beta})E_{\Delta}(H)
-c_{2}\frac g \beta E_{\Delta}(H) \\
&\geq (\frac \beta 2 -\frac{cg}{\beta}) E_{\Delta}(H)
\end{split}\end{align*}
for some $c>0$ independent of 
$\Delta$, $g$ and $\beta$. 
\qed

\medskip

\textit{Proof of theorem \ref{thm5}} 
Theorem \ref{thm5} is a consequence of proposition \ref{prop5.2} and 
the Mourre theory. Actually it only remains to verify the 
applicability of this theory. This means that we have to verify that 
$[A,H]$ and $[A,[A,H]]$ are $H$ bounded. From \eqref{H0A} we deduce 
that $[A,H_{0}]$ is  $H_{0}$ bounded. For the second commutator a simple 
calculus gives
\begin{align*}\begin{split}
    [A,[A,H_{0}]]\psi &=\Big[\sum_{\ep=\pm}d\Gamma_{1\ep}\left(\frac{p^2m_{1}^2}
    {(p^2+m_{1}^2)^{3/2}}\right)
+d\Gamma_{2\ep}(\va{p})+d\Gamma_{3\ep}(\va{p})\\
&+d\Gamma_{4\ep}\left(\frac{p^2m_{4}^2}{(p^2+m_{4}^2)^{3/2}}\right)\Big]
\psi
\end{split}\end{align*}
for $\psi \in \F_{\mathrm{fin}}$.
Thus $[A,[A,H_{0}]]$ is $H_{0}$ bounded.\\
We have already noted that $[A,H_{I}]$ is $H_{0}$ bounded as soon as
$a_{j}G_{\ep\ep'}\in L^2$ for each 
$j=1,2,3,4$ and for each $\ep\neq \ep'$. The computation of the commutator of 
$A$ with the expression of
$[A,H_{I}]$ given by \eqref{HIA}
shows that $[A,[A,H_{I}]]$ is $H_{0}$ bounded as soon as
$a_{j}a_{j}G_{\ep\ep'}\in L^2$ for each 
$j=1,2,3,4$ and for each $\ep\neq \ep'$. These conditions on 
$G_{\ep\ep'}$ are satisfied when $p_{i}\cdot \nabla_{p_{i}}G_{\ep\ep'}$ and 
    $p_{i}^2\Delta_{p_{i}}G_{\ep\ep'}$ are all in 
    $L^2(\Sigma_{1}\times\Sigma_{2}\times\Sigma_{2}\times\Sigma_{1})$ for $i=1,2,3,4$ 
    and $\ep\neq \ep'$.
    \qed

\section{Other examples}

The main other examples of the Fermi-weak interactions are the beta 
decay of the neutron and of the quarks $u$ and $d$. Let us consider 
the decay of the quark $d$. This decay involves four species of 
particles and antiparticles: the quarks $u$ and $d$ and their 
antiparticles $\bar u$ and $\bar d$, the electron $e^-$ and the 
positron $e^+$, the neutrino $\nu_{e}$ and its antineutrino 
$\bar\nu_{e}$ (see \cite{Wei, Gre}). The Fock space is the fermionic Fock space associated 
to these four species of particles and the interaction is given by 
\bas
H_{I}=&\int d\xi_{1}d\xi_{2}d\xi_{3}d\xi_{4}
\ J(\xi_{1},\xi_{2},\xi_{3},\xi_{4})
\ b^\star_{1,+}(\xi_{1})b^\star_{2,-}(\xi_{2})
b^\star_{3,+}(\xi_{3})b_{4,+}(\xi_{4})\\
&+\int d\xi_{1}d\xi_{2}d\xi_{3}d\xi_{4}\
\overline{J(\xi_{1},\xi_{2},\xi_{3},\xi_{4})}
\ b^\star_{4,+}(\xi_{4})b_{3,+}(\xi_{3})
b_{2,-}(\xi_{2})b_{1,+}(\xi_{1}).
\end{split}\end{align}
Here the particles and antiparticles 1 are the electrons and the positrons, 
the particles and antiparticles 2 are the  neutrinos $\nu_{e}$ and 
$\bar\nu_{e}$, the particles and antiparticles 3 are the quarks $u$ 
and $\bar u$ and, finally, the particles and antiparticles 4 are the 
quarks $d$ and $\bar d$.\\
Obviously  theorems \ref{thm4} and \ref{thm5} remains valid for the 
associated Hamiltonian under appropriate conditions on the kernel 
$J$.\\
We can also consider the decay of the massive bosons $W^\pm$ into 
electrons, positrons and neutrinos $\nu_{e}$ and $\bar \nu_{e}$ (see \cite{Wei, Gre}). The 
Fock space is the tensor product of the fermionic
Fock space associated to the electrons, the positrons and the 
neutrinos $\nu_{e}$ and $\bar \nu_{e}$ and of the bosonic Fock space 
associated to a massive boson of spin 1. The interaction is then 
given by
\bas
H_{I}=&\sum_{\ep\neq\ep'}\int d\xi_{1}d\xi_{2}d\xi_{3}\
K_{\ep,\ep'}(\xi_{1},\xi_{2},\xi_{3})\
 b^\star_{1,\epsilon}(\xi_{1})b^\star_{2,\epsilon'}(\xi_{2})
a_{3,\epsilon}(\xi_{3})\\
&+\sum_{\ep\neq\ep'}\int d\xi_{1}d\xi_{2}d\xi_{3}\
\overline{K_{\ep,\ep'}(\xi_{1},\xi_{2},\xi_{3})}\
a^\star_{3,\epsilon}(\xi_{3})
b_{2,\epsilon'}(\xi_{2})b_{1,\epsilon}(\xi_{1}).
\end{split}\end{align}
Here the particles and antiparticles 1 are the electrons and the positrons, 
the particles and antiparticles 2 are the  neutrinos $\nu_{e}$ and 
$\bar\nu_{e}$, and $a_{+}(\xi_{3})$ (resp. $a_{-}(\xi_{3})$) is the 
annihilation operator for the meson $W^-$ (resp. $W^+$).\\
Once again theorems \ref{thm4} and \ref{thm5} remains valid for the 
associated Hamiltonian under appropriate conditions on the kernels 
$K_{\ep,\ep'}$.\\
One could also give a mathematical model for the decay of the massive 
boson $Z^0$.


\bigskip

{\bf Laurent AMOUR}

{\it Laboratoire de Math\'ematiques EDPPM, UMR-CNRS 6056,

Universit\'e de Reims,

Moulin de la Housse - BP 1039, 

51687 REIMS Cedex 2, France. 

\smallskip
E-mail: {\tt laurent.amour@univ-reims.fr}
}

\medskip 
 
{\bf Beno\^{i}t Gr\'ebert}

{\it Laboratoire de Math\'ematique Jean Leray UMR 6629,

Universit\'e de Nantes,

2, rue de la Houssini\`ere,

44322 Nantes Cedex 3, France

\smallskip
E-mail: {\tt benoit.grebert@univ-nantes.fr}
}

\medskip

{\bf Jean-Claude GUILLOT}

{\it CMAP, Ecole polytechnique, CNRS,

Route de Saclay

91128 Palaiseau, France.

\smallskip
E-mail: {\tt guillot@cmapx.polytechnique.fr}
}

\end{document}